\documentclass[runningheads]{llncs}
\usepackage{graphicx}
\usepackage[utf8]{inputenc}
\usepackage{amsmath}
\usepackage[lambda,landau,operators,probability,sets,logic,advantage,complexity,asymptotics,notions,mm,primitives,ff,adversary,keys]{cryptocode}
\usepackage{hyperref}
\usepackage{xspace}
\usepackage{bm}
\usepackage{dashrule}
\usepackage{physics}
\usepackage{amsmath}
\usepackage{bigstrut}
\usepackage{tikz}
\usepackage{mathdots}
\usepackage{yhmath}
\usepackage{cancel}
\usepackage{color}
\usepackage{siunitx}
\usepackage{array}
\usepackage{multirow}
\usepackage{amssymb}
\usepackage{gensymb}
\usepackage{tabularx}
\usepackage{extarrows}
\usepackage{booktabs}
\usepackage{listings}
\usepackage{xcolor}
\usepackage{appendix}
\usepackage{wrapfig}

\usetikzlibrary{fadings}
\usetikzlibrary{patterns}
\usetikzlibrary{shadows.blur}
\usetikzlibrary{shapes}

\newcommand{\point}[1]{\par\smallskip{\noindent\textbf{#1.}~}}

\newcommand{\MeanEndToEndExecutionTime}{48.4~seconds\xspace}
\newcommand{\GasUsedForVerification}{191,687~gas\xspace}
\newcommand{\GasUsedForPublicParams}{281,715~gas\xspace}
\newcommand{\GasUsedTotal}{473,402~gas\xspace}

\newcommand{\PerUserCost}{\texttt{$\sim$}\$0.15\xspace}
\newcommand{\MultipleFromTrading}{2.4}
\newcommand{\AvgMultipleFromTrading}{1.4}
\newcommand{\tool}{\textsc{ChainBot}\xspace}
\newcommand{\etal}{\emph{et~al.}\xspace}
\newcommand{\UserInvestment}{1,000\xspace}
\newcommand{\MonthlyGasCost}{3\xspace}
\newcommand{\MonthlyEarning}{105\xspace}

\usepackage{listings, xcolor}

\definecolor{verylightgray}{rgb}{.97,.97,.97}

\lstdefinelanguage{Solidity}{
	keywords=[1]{anonymous, assembly, assert, balance, break, call, callcode, case, catch, class, constant, continue, constructor, contract, debugger, default, delegatecall, delete, do, else, emit, event, experimental, export, external, false, finally, for, function, gas, if, implements, import, in, indexed, instanceof, interface, internal, is, length, library, log0, log1, log2, log3, log4, memory, modifier, new, payable, pragma, private, protected, public, pure, push, require, return, returns, revert, selfdestruct, send, solidity, storage, struct, suicide, super, switch, then, this, throw, transfer, true, try, typeof, using, value, view, while, with, addmod, ecrecover, keccak256, mulmod, ripemd160, sha256, sha3}, 
	keywordstyle=[1]\color{blue}\bfseries,
	keywords=[2]{address, bool, byte, bytes, bytes1, bytes2, bytes3, bytes4, bytes5, bytes6, bytes7, bytes8, bytes9, bytes10, bytes11, bytes12, bytes13, bytes14, bytes15, bytes16, bytes17, bytes18, bytes19, bytes20, bytes21, bytes22, bytes23, bytes24, bytes25, bytes26, bytes27, bytes28, bytes29, bytes30, bytes31, bytes32, enum, int, int8, int16, int24, int32, int40, int48, int56, int64, int72, int80, int88, int96, int104, int112, int120, int128, int136, int144, int152, int160, int168, int176, int184, int192, int200, int208, int216, int224, int232, int240, int248, int256, mapping, string, uint, uint8, uint16, uint24, uint32, uint40, uint48, uint56, uint64, uint72, uint80, uint88, uint96, uint104, uint112, uint120, uint128, uint136, uint144, uint152, uint160, uint168, uint176, uint184, uint192, uint200, uint208, uint216, uint224, uint232, uint240, uint248, uint256, var, void, ether, finney, szabo, wei, days, hours, minutes, seconds, weeks, years},	
	keywordstyle=[2]\color{teal}\bfseries,
	keywords=[3]{block, blockhash, coinbase, difficulty, gaslimit, number, timestamp, msg, data, gas, sender, sig, value, now, tx, gasprice, origin},	
	keywordstyle=[3]\color{violet}\bfseries,
	identifierstyle=\color{black},
	sensitive=false,
	comment=[l]{//},
	morecomment=[s]{/*}{*/},
	commentstyle=\color{gray}\ttfamily,
	stringstyle=\color{red}\ttfamily,
	morestring=[b]',
	morestring=[b]"
}

\usepackage[color=pink]{todonotes}

\hypersetup{colorlinks=true,linkbordercolor=blue,linkcolor=blue,pdfborderstyle={/S/U/W 1}}

\usepackage{xcolor}
\usepackage{colortbl}
\usepackage{colortbl}
\definecolor{yellow}{rgb}{1,1,0.6}
\begin{document}\sloppy
\title{Towards Private On-Chain Algorithmic Trading}
%

\author{Ceren Kocaoğullar\inst{1}\and
Arthur Gervais\inst{2} \and
Benjamin Livshits\inst{2}}

\authorrunning{C. Kocaoğullar et al.}
%
\institute{University of Cambridge \and
Imperial College London}
\maketitle              
\begin{abstract}

While quantitative automation related to trading crypto-assets such as ERC-20 tokens has become relatively commonplace, with services such as 3Commas and Shrimpy offering user-friendly web-driven services for even the average crypto trader, not the mention the specialist, we have not yet seen the emergence of on-chain trading as a phenomenon. We hypothesize that just like  decentralized exchanges~(DEXes) that by now are by some measures more popular than traditional exchanges, process in the space of decentralized finance~(DeFi) may enable attractive online trading automation options.

In this paper we present \tool, an approach for creating algorithmic trading bots with the help of blockchain technology. We show how to partition the algorithmic computation into on- and off-chain components in a way that provides a measure of end-to-end integrity, while preserving the algorithmic ``secret sauce''. The end result is a system where an end-user can sign-up for the services of a trading bot, with trust established via on-chain publicly readable contracts. Our system is enabled with a careful use of algorithm partitioning and zero-knowledge proofs together with standard smart contracts available on most modern blockchains. 

Our approach offers more transparent access to liquidity and better censorship-resistance compared to traditional off-chain trading approaches both for crypto- and more traditional assets. We show that the end-to-end costs and execution times are affordable for regular use and that gas costs can be successfully amortized because trading strategies are executed on behalf of large user pools. Finally, we show that with modern layer-2~(L2) technologies, trades issued by the trading bot can be kept private, which means that algorithmic parameters are difficult to recover by a chain observer.

With \tool, we develop a sample trading bot and train it on historical data, resulting in returns that are up to~$\MultipleFromTrading\times$ and on average~$\AvgMultipleFromTrading\times$ the buy-and-hold strategy, which we use as our baseline. 
Our measurements show that across~1000 runs, the end-to-end average execution time for our system is~\MeanEndToEndExecutionTime. 
We demonstrate that the frequency of trading does not significantly affect the rate of return and Sharpe ratio, which indicates that we do not have to trade at every block, thereby significantly saving in terms of gas fees.
In our implementation, a user who invests~\$\UserInvestment would earn~\$\MonthlyEarning, and spend~\$\MonthlyGasCost on gas; assuming a user pool of~1,000 subscribers. 

\keywords{Blockchain privacy, Algorithmic trading, Decentralized finance}
\end{abstract}

\section{Introduction}
\label{sec:intro}

Electronic financial markets utilise trading algorithms to automate their trading strategies, processing financial data, news and such to make predictions about the state of the market. The rise of Decentralized Finance (DeFi) have paved the way for algorithmic trading in cryptomarkets. With assets as well as transactions having presence on distributed, decentralized and public blockchains, crypto-trading bots have the potential to provide security and integrity properties unparalleled in traditional settings. 

Trading on-chain with a crypto-trading bot brings unique benefits. For instance, on-chain liquidity provides transparency in terms of costs and availability. Moreover, centralized systems fall short in privacy and censor-resistance, a recent example being the regulatory pressures against the prominent centralized cryptomarket Binance \cite{binancereuters}. 

\begin{figure}[tb]
\lstset{escapeinside={(*@}{@*)}}
\begin{lstlisting}[language=Solidity]
function algoTrade(uint numPeriods, uint periodSize, uint upperPerc, uint lowerPerc, uint amount) public {

   require(msg.sender == admin); (*@\label{lst:require}@*)
   uint currentTime = block.timestamp;
   uint price = Oracle.getPriceAt(currentTime);
   uint movingAvg = SMA(currentTime, numPeriods, periodSize);
   
   if (price > (movingAvg / 100) * (100 + upperPerc)) 
      DEX.swap(token1, token0, amount);
   else if (price < (movingAvg / 100) * (100 - lowerPerc))
      DEX.swap(token0, token1, amount);
}
\end{lstlisting}
\label{fig:strawman}
\caption{A strawman on-chain trading algorithm example written in Solidity, which uses the Simple Moving Average~(SMA) indicator.}
\end{figure}


This paper lays the foundation for private on-chain algorithmic trading, an approach we call \tool. 
On-chain trading has the advantage of full transparency favored by DeFi advocates and users, however, transparency kills the competitive advantage a trading algorithm might have over its competitors. Running the algorithms off-chain, as crypto-trading bots 3Commas \cite{3commas}, Shrimpy \cite{shrimpy} and others do, may enable privacy; however, without the transparency and integrity properties of on-chain execution, users have to trust the off-chain bots with executing trading algorithms and handling user funds correctly. The focus of this work is an approach that combines on-chain and off-chain computation to create on-chain trading algorithms that are trust-minimized and transparent yet able to execute in a way that does not destroy their competitive advantage due to privacy loss.

\point{Strawman example of on-chain trading}
Consider a simple trading algorithm that estimates the price by calculating price averages within a moving history, namely the Simple Moving Average~(SMA), and buys (or sells) if the actual price is lower (or higher) than a determined lower (or upper) bound~\cite{pole2011statistical}. The Solidity function presented in Figure~\ref{fig:strawman} implements this algorithm for on-chain execution, calculating
\begin{enumerate}
\item the SMA (\texttt{movingAvg}) starting from \texttt{currentTime} with a window of \texttt{numPeriods}-many \texttt{periodSize}-second intervals; 
\item lower and upper bounds \texttt{lowerPerc} and \texttt{upperPerc} below and above the \texttt{movingAvg}, respectively. 
\end{enumerate}
Calling this function deployed on a smart contract will execute automated trades on-chain, however, the trading algorithm, parameters it takes and the resulting trades will be public information. Therefore, even though line~\ref{lst:require} of Figure~\ref{fig:strawman} restricts function calls to someone assigned as the administrator, anyone can copy this function and deploy it to another smart contract or watch the trades of this contract to mirror its actions. This simple example highlights a fundamental problem in DeFi: if the code can be read, it can be cloned as well. While some DeFi protocols have taken to obfuscating Solidity code and putting obfuscated EVM code into their contracts, cloning generally remains possible. \tool suggests a fundamentally better approach. 

\subsection{Contributions}
This paper makes the following contributions:

\begin{itemize}
\item \textbf{Combining on-/off-chain.}
In this paper, we propose a hybrid on-/off-chain approach we call \tool that combines the benefits of on-chain integrity and transparency with the privacy benefits of off-chain execution. Off-chain integrity is maintained via zero-knowledge proofs based on Zokrates~\cite{eberhardt2018zokrates}. 

\item \textbf{Parametrized algorithms.}
We show how to use the hybrid execution strategy to build automatic crypto-trading bots, where the (public) code of the bot is parametrized by (confidential) parameters derived by training on historical data. 

\item \textbf{Confidential trading.}
We further show how to run these on-/off-chain bots in a way that uses a private L2-based decentralized exchange and on-chain oracles for price data. 

\item \textbf{Experimental evaluation.} 
We train our selected parametrized trading algorithm on historical ETH:USDC price data and illustrate that training can yield up to~$\MultipleFromTrading\times$ and on average ~$\AvgMultipleFromTrading\times$ returns compared to the buy-and-hold strategy, effectively indicating that the \tool approach is viable for algorithmic traders. The experimental evaluation of our implemented \tool show that end-to-end execution of each trade takes~\MeanEndToEndExecutionTime on average. We demonstrate that trading frequency does not effect returns significantly, signalling reduced gas fees. The cost analysis exhibits that the system is able to sustain continuous trading with a per-trade amortized cost of~\PerUserCost per user, assuming a user base of~1,000 people. 

\end{itemize}
\section{Background}
\label{sec:background}

This section discusses the trade-offs of designing on- and off-chain systems that work together in Section~\ref{sec:on-off-chain}; Section~\ref{sec:blockchain-privacy} talks about privacy-preserving techniques for public blockchains. Lastly, Section~\ref{sec:algo-trading} discusses the basics of algorithmic trading.

\subsection{On-Chain/Off-Chain Trade-offs}
\label{sec:on-off-chain}
Blockchain technology has several inherent security properties such as consistency and availability enabled by decentralization; tamper-resistance through storing chains of hashes; and resistance to double-spending by using cryptographic signatures~\cite{zhang2019security}. On the other hand, having a permissionless consensus structure and propagating every state change throughout the network leads to low transaction throughput, existence of transaction fees and lack of privacy. 

\emph{Layer-2 (L2)} protocols built on top of (\emph{layer-1/L1}) blockchains can improve scalability by carrying out transactions off-chain and submitting a smaller number of assertions or proofs to the blockchain than the number of transactions~\cite{gudgeon2020sok}. Notable L2 protocols include Plasma~\cite{poon2017plasma}, which uses side-chains, and Lightning Network~\cite{poon2016lightning}, which has state channels. Another notable L2 approach is using optimistic or ZK-rollups, where multiple off-chain transactions are batched in a fraud or zero-knowledge proof, respectively, and submitted to the chain \cite{rollups}.

Despite increasing throughput and decreasing transaction costs, these solutions do not fully preserve the privacy, as they keep the transaction data available to at least some of the participants or reveal it in case of dispute. Validium-style L2 solutions such as StarkEx~\cite{starkex} (the basis for DiversiFi~\cite{deversifi} exchange) keep both computation and data off-chain. However, this leads to a potential \emph{data availability problem}, preventing those who do not have access to the transaction data from reconstructing the state. The current solution to this issue is passing the responsibility of making sure that data is available to a Data Availability Committee~(DAC), falling short in providing full privacy.

\subsection{Blockchain Privacy}
\label{sec:blockchain-privacy}
Gudgeon~\etal provide a survey of L2 blockchain protocols, so of which focus on privacy, mainly based on submitting assertions or proofs to the chain~\cite{gudgeon2020sok}. 
\point{Off-chain computation} TrueBit project provides an off-chain computation layer where computational tasks can be outsourced in a privacy-preserving way, however, disputed data should be made public since \emph{fraud proofs} are used~\cite{Teutsch2019}. {zkay}~\cite{Report2020,Steffen2019} a high-level programming language sidesteps this issue by solely depending on cryptographic primitives; on the other hand, the cost of verifying zero-knowledge proofs on-chain may be high.

One way of achieving privacy with off-chain computation is by mixing cryptography and incentivization, as in Hawk~\cite{Kosba2016}, Arbitrum~\cite{Kalodner2018} and Ekiden~\cite{Cheng2019}. Nevertheless, this approach comes at the cost of losing the intrinsic trustlessness of blockchain: Hawk and Arbitrum have trusted managers and Ekiden depends on trusted hardware. There have been efforts to replace trusted entities with secure multi-party computation (MPC) and similar mechanisms, such as Raziel~\cite{Sanchez2018}, zkHawk~\cite{Banerjee2021}, Enigma~\cite{Zyskind2019}, and ZEXE~\cite{Bowe2020}; however, carrying out computation on encrypted data is a heavyweight and costly task, as recognised by these works. 

\point{Privacy-focused L1s}
The prominent privacy-preserving payment protocols such as Zerocash~\cite{sasson2014zerocash} and cryptocurrencies like Zcash~\cite{hopwood2016zcash}, Solidus~\cite{cecchetti2017solidus}, Monero~\cite{monero} and several others~\cite{danezis2013pinocchio,van2013cryptonote,jivanyan2019lelantus,bunz2020zether,fauzi2019quisquis} offer confidential transactions but cannot be used for general computation. Similarly, non-privacy preserving blockchains have seen the adoption of add-on mixers~\cite{le2020amr}. Lastly, blockchains designed with privacy considerations, such as Dusk~\cite{Maharramov2019}, can enable privacy, but having a separate chain might cause complicate integration with existing platforms and features. In recent years, popular privacy-focused chains have been subjected to a number of deanonimization attacks~\cite{beres2020blockchain,tracingYoussaf,chervinski2019floodxmr,moneroAttack}. 

\subsection{Algorithmic Trading of Cryptocurrencies}
\label{sec:algo-trading}
Digitalization of data and operations in traditional markets have enabled the usage of computer algorithms for making trading decisions~\cite{hendershott2009algorithmic}. 
In~2011, over~73\% of equity volume in the U.S. were handled with algorithmic trading~\cite{treleaven2013algorithmic}. Despite being a highly influential portfolio and asset management tool, the inherent competition, scantiness of expertise, and high profitability cause algorithmic trading to be an opaque subject~\cite{chan2021quantitative,chan2013algorithmic,kim2010electronic,pole2011statistical,treleaven2013algorithmic}.

\point{Algorithmic trading roadmap}
Determining a trading algorithm is the first step of algorithmic trading; Some of the basic traditional algorithms include Volume-Weighted Average Price~(VWAP), Time-Weighted Average Price~(TWAP), implementation shortfall, and arrival price~\cite{kim2010electronic}. Once a trading algorithm is chosen or created, it should be \emph{backtested}, i.e. run on historical data to evaluate its performance~\cite{chan2021quantitative,chan2013algorithmic}. After a successful backtest, real-world trading can proceed. 
This is a multi-step process: First, real-time or historic financial, economic or social/news data, should be collected, prepared, and analyzed. The selected algorithm then uses the data to make a trade decision, the trade is executed, and lastly, analyzed~\cite{treleaven2013algorithmic}.

\point{Blockchain automated trading}
One of DeFi's objectives, when compared to traditional finance, is to foster financial transparency and accountability~\cite{qin2021cefi}, while flash loans provide novel atomic lending contracts granting instantaneously billions of USD in debt~\cite{qin2020attacking}. The assets, as well as the operations being digital in blockchain, DeFi has facilitated handling many tasks related to governance, issuance, and order matching algorithmically~\cite{Calcaterra2021}. The same trend is observed in trading and portfolio management: several studies have shown that machine learning can be used for Bitcoin price prediction~\cite{shahandzhang} or directly outputting the portfolio vector~\cite{jiang2017cryptocurrency} to automate Bitcoin trading and execute high-frequency trading~\cite{vo2020high}. A recent study has shown that high-frequency trading can occur in adversarial form, taking advantage of the inherently deterministic Automated-Market Maker DEXes~\cite{Zhou2020} to exploit blockchain extractable value~\cite{qin2021quantifying}. Regarding high-frequency trading, recurring price discrepancies among cryptoexchanges generate arbitrage opportunities~\cite{makarov2020trading}, enabling cross-market statistical arbitrage~\cite{pole2011statistical,zhou2021a2mm} and automated tools generate profitable transactions finding arbitrage and attack vectors~\cite{zhou2021just}. 

\point{Crypto-bots}
Besides these academic efforts, commercial and community-driven portfolio management protocols include web-based trading bots that automate simple trading strategies~\cite{3commas,shrimpy}, savings accounts running simultaneous investment strategies~\cite{yearnfinance}, rebalancing token sets~\cite{tokensets}. Another prominent category of automated traders is \emph{social trading platforms} that copy the strategies of experienced traders~\cite{3commas,coinmatics,shrimpy,tradelize}. 
\section{Overview}
\label{sec:overview}

This section presents an overview of the concepts and components that form \tool. Section~\ref{sec:private-trading} presents a blueprint for private on-chain algorithmic trading. Section~\ref{sec:system-components} describes the on- and off-chain components of the system and their roles. Lastly, Section \ref{sec:establishing-integrity} discusses how integrity is preserved end-to-end throughout on-chain/off-chain interactions and data exchanges.

\subsection{Private On-Chain Algorithmic Trading}
\label{sec:private-trading}
Commercial trading bots that execute trades on behalf of users in exchange for a fee or commission require user trust to a significant degree, therefore incompatible with the DeFi's convergence towards minimal trust. Notably, the users of an off-chain trading bot cannot verify that the bot uses a legitimate trading algorithm to execute trades, even in the case that an algorithm is disclosed to the users. Moreover, the users have to trust that the bot will be managing their funds with integrity and will not steal, withhold, or lose them. 

Executing the trading algorithm on-chain eliminates the need for trust, since users of a fully on-chain bot can audit the trading algorithm and make sure that it is being correctly executed. Moreover, the user funds can be stored in a smart contract safely and the users can have full control over their money. On the other hand, with the algorithm open to the public, the bot loses its competitive advantage, as indicated by the existence of \emph{social trading platforms}~\cite{3commas,coinmatics,shrimpy,tradelize} that mirror the strategies of well-performing traders. 

We suggest that an optimal mix of off-chain and on-chain computation can enable algorithmic trading with integrity, minimal trust and increased transparency while conserving competitive advantage. The core idea is having the skeleton and public parameters of a parametrized trading algorithm on-chain, while keeping the parameters critical to the algorithm's performance private by storing them off-chain. These private parameters can be periodically optimized through training the trading algorithm off-chain, adapting to changing market conditions. Secrecy of private parameters are preserved through running the trading algorithm off-chain and using zero-knowledge proofs to verify on-chain that the algorithm is run correctly. Once a proof is verified, the trades should be executed privately to protect private parameters from being reverse engineered.

\begin{figure*}[tbp]
\centering\scriptsize
\begin{tabular}{p{.33\textwidth}p{.33\textwidth}p{.33\textwidth}}
\toprule
     \bf On-chain bot &  \bf Off-chain bot & \bf Proof system \\ \midrule
     
  Obtaining price information from the on-chain price oracle &  \cellcolor{yellow} Performing algorithmic training to find the optimal private parameters & \cellcolor{yellow} Containing the Zokrates program that bears the trading logic to be proven\\
  \cellcolor{yellow} Calculating public parameters of the trading algorithm, such as financial indicators & \cellcolor{yellow} Making trade decisions based on the trading algorithm using public and private parameters & Generating the verifier contract from the Zokrates code and deploying it to the chain \\
  Checking the correctness of public parameters including the price used in each trade decision & Trading on the privacy-enhancing DEX once a proof has been verified &  Creating a zero-knowledge proof of each trading decision, without revealing any information about private parameters\\
\bottomrule
\end{tabular}
\caption{Functionalities that each \tool unit is responsible for. Those that need to be customized according to the trading algorithm are colored, generic functionalities are uncolored.}
\label{fig:system-component-roles}
\end{figure*}

\subsection{System Components}
\label{sec:system-components}
\tool consists of multiple on-chain and off-chain elements categorized under three main units based on their roles in the system: (1)~on-chain bot, (2)~off-chain bot and (3)~proof system. In addition to these internal units, the system requires interaction with two external entities: (1) An L2 privacy-enhancing DEX such as DeversiFi~\cite{deversifi} (see Appendix \ref{sec:diversifi} for implementation details of Deversifi in our system) and (2) an on-chain oracle, such as Uniswap~\cite{Adams2021} or ChainLink~\cite{Ellis2017}. Data used in feeding the algorithms can be simple as price or more complex as orderbook data, so long as the data provider is an on-chain oracle.

The main units of \tool are described below. Functionalities of each unit is listed in Figure \ref{fig:system-component-roles}, classified as generic and algorithm-dependent custom functionalities to provide direction for the developer who wants to deploy their own \tool. 
\begin{itemize}
    \item \textbf{On-chain bot} is a smart contract that is responsible for obtaining and calculating public parameters (price, financial indicators, etc.) and ensuring that they are used correctly in trade decisions.
    \item \textbf{Off-chain bot} trains the algorithm (see Section \ref{sec:eval-algorithmic-training} and Appendix \ref{sec:appendix-training}), stores private parameters, makes trade decisions and executes trades. 
    \item \textbf{Proof system} consists of an on-chain verifier and off-chain proof generator. It keeps the private parameters hidden from all other than the off-chain bot while ensuring the correctness of trade decisions. Zokrates\footnote{https://zokrates.github.io/}\cite{eberhardt2018zokrates} or any zero-knowledge proof system that allows on-chain verification can be used here. Appendix \ref{sec:zokrates} discusses the implementation details of Zokrates in our system.
\end{itemize}

\begin{figure*}[h!]
\centering
\resizebox{\textwidth}{!}{%
\input{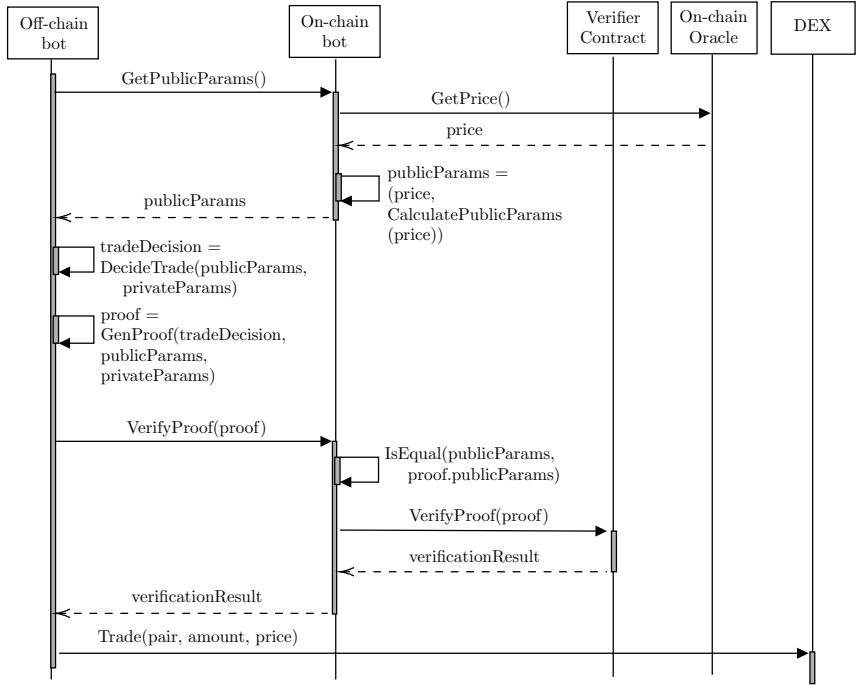}
}
\caption{Trading bot architecture as a sequence diagram.}
\label{fig:sequence-diag}
\end{figure*}
Each trade is a sequence of events and interactions between these three units as well as the two external entities, namely the privacy-enhancing DEX and the on-chain oracle. Figure \ref{fig:sequence-diag} describes the detailed end-to-end algorithmic trade process that can be summarized as follows: Off-chain bot requests public parameters from the on-chain bot, which obtains the price data from the price oracle and calculates the other public parameters if necessary. Once receiving the public parameters, off-chain bot runs the algorithm with public and private parameters to make a trading decision and creates a proof of it. The proof is verified on-chain by the verifier contract, while the on-chain bot checks the correctness of public parameters. Once a proof is verified, the off-chain bot executes the trade on the privacy-enhanced DEX.

It is important to note that the diagram does not include user subscription stages, as they are not fundamental to the trading process. We outline the necessary conditions for user subscription as follows:
\begin{itemize}
    \item Each user subscribes to \tool by sending their funds to the on-chain bot.
    \item On-chain bot records the amount invested by each user.
    \item In each trade, on-chain bot uses the pooled user funds for trading and paying execution fees.
    \item Once a trading epoch is complete, the on-chain bot pays each user their share of returns. 
\end{itemize}

\newcommand{\assumptionN}[1]{\textsf{{\color{blue} #1}}\xspace}

\subsection{Algorithmic Training}\label{sec:overview-algo-training}
We implement a simple algorithmic training method using grid search to demonstrate the process of creating parameters that can bring competitive advantage to the bot owner, therefore she may want to keep secret. This method aims to produce a trading algorithm that outperforms arguably the most basic of trading strategies~---~buy-and-hold. 

\point{The trading algorithm}
We adopt a trading algorithm that uses Bollinger Bands~(BBs)~\cite{bollinger1992using}, a technical analysis tool that considers the ``envelope'' around a simple moving average using a number of standard deviations both above (\emph{upper Bollinger Band}) and below (\emph{lower Bollinger Band}) (see Appendix \ref{sec:appendix-training}, Equation \ref{eq:bollinger}). Our trading algorithm signals a buy (sell) if the current price is within a threshold percentage range below (above) of the lower (upper) Bollinger Band (see Appendix \ref{sec:appendix-training}, Equation \ref{eq:trading-algo}).

There are two main reasons behind our choice of algorithm, which we refer to as \emph{Bollinger algorithm}: 
\begin{enumerate}
    \item 
    Through backtesting, we have observed that this trading strategy is simple yet powerful enough to make profit when buy and hold is an unprofitable strategy. 
    \item 
    It is suitable for parametrization as described below.
\end{enumerate}

\point{Period selection}
To determine training and testing periods, we find all 30-day periods within a date range where buy-and-hold was an unprofitable strategy. The first half of these periods are allocated for training and the rest for testing. It is important to note that changing how often is a trade is executed, i.e. the trading period, may also change the trading and testing periods. 

\point{Training}
Algorithmic training is based on parametrization of the algorithm and optimizing those parameters. We parametrize Bollinger algorithm with four parameters: 
\begin{itemize}
    \item moving average period size~(\emph{N}), 
    \item number of standard deviations~(\emph{D}), 
    \item percentage threshold that the price should be from the upper Bollinger band to sell~(\emph{U}), and 
    \item the lower Bollinger band to buy~(\emph{L}). 
\end{itemize}
We train the algorithm through grid search, feeding it the minimum, maximum, and mean values of each parameter. Finally, we rank the parameter configurations based on performance.

\point{Testing} Once we determine the top-performing parameter configurations, we run the trading algorithm with each configuration throughout the testing periods and record the relative returns to buy-and-hold strategy. 

\subsection{Trust Assumptions}
\tool makes the following trust assumptions:

\point{\assumptionN{A1}: Blockchain security}
We ignore the ramifications of blockchain forks~\cite{gervais2016security}, and assume that the majority of consensus participants are honest. We further ignore bugs and security vulnerabilities in smart contracts.

\point{\assumptionN{A2}: Validity of the trusted setup} 
In zero-knowledge proof systems using zk-SNARKs, verification and proving keys are created based on a \emph{trusted setup} phase. Since multiple users are interested in the correctness of the proofs in our setting, a so-called \emph{ceremony} that likely involves MPC should be used for the trusted setup~\cite{zcash-ceremony,Report2020}.

\point{\assumptionN{A3}: Off-chain bot fund integrity} 
In \tool, we assume that the off-chain bot manages the user funds with integrity, i.e. it does not steal or withhold but only uses assets for trading. We also assume that the bot performs algorithmic training honestly and favorably, feeding the trained parameters properly into the proof. It is important to highlight that these assumptions do not include any reliance to the off-chain bot for executing the trading algorithm correctly. Moreover, although an off-chain bot is capable of stealing or withholding user funds, unlike in an off-chain trading bot, users can verify the flow of their funds by querying the blockchain ledger. One way to ensure that in practical terms might be to use permissioned trading APIs of centralized exchanges such as Binance, etc. for off-chain bot operations specifically designed to address the bot trading scenario. 

\point{\assumptionN{A4}: DEX integrity}
In terms of the external entities that \tool interacts with, we assume that the privacy-enhancing DEX enables a method for ensuring that trades are executed correctly based on the verified financial decisions. For instance, the DEX can provide a transaction hash that can be checked against a Merkle root by anyone and used for verifying the mentioned information through a zero-knowledge proof. 
We further assume that the DEX either provides full privacy for trades or if it shares any trade data with any parties (such as the DAC in DeversiFi --- see Appendix \ref{sec:diversifi}), they are impartial and honest.

\point{\assumptionN{A5}: DEX privacy}
We assume that the DEX is able to make trades in a way that does not reveal trade types (i.e. asset pair), sources and destinations, as well as amounts to blockchain observers. 

\point{\assumptionN{A6}: Oracle integrity}
We assume that the on-chain price oracle provide accurate data. Attacks against oracles are outside of this paper's scope~\cite{eskandari2021sok}.

\subsection{Threat Analysis}
\label{sec:establishing-integrity}

\tool is designed to prevent two types of threats: end-to-end integrity threats to the trading bot and theft of the trading algorithm. 

\point{End-to-end trading integrity} 
More specifically, this entails the following:

\begin{enumerate}
    \item verifying that the correct public parameter values are used in trade decisions, enabled by the proof system's ability to include some parameters openly in the proof, allowing the on-chain bot to verify that the proof indeed includes the correct public parameters provided (see Section~\ref{sec:zokrates}); this uses \assumptionN{A1} and~\assumptionN{A2}
    \item guaranteeing that the private parameters are included correctly in the algorithm, guaranteed by the zero-knowledge proof; this uses \assumptionN{A1};
    \item ensuring that trading decisions are made based on correct asset pricing as per \assumptionN{A6};
    \item ensuring that the trades are executed based on the trading decision, combined with the assumption that the DEX provides a way to verify trades in a way that does not damage user privacy; this uses \assumptionN{A3}, \assumptionN{A4}, and \assumptionN{A5}. 
\end{enumerate}

\point{Algorithmic privacy} Algorithmic privacy stems from the fact that in \tool we use parametrized algorithms where parameters are kept off-chain and therefore cannot be easily stolen or replicated. Further, our reliance on a private DEX makes reverse-engineering these parameters all but impossible. This uses~\assumptionN{A3} and \assumptionN{A5}.

\begin{figure*}[tb]
\centering\small
\resizebox{\textwidth}{!}{%
\begin{tabular}{|l|r|rrrr|}
\hline
\bf Ranking method                  & \bf Parameter config. & \bf Max.              & \bf Min.            & \bf Mean             & \bf Standard dev.     \bigstrut\\ 
\hline
\multirow{6}{*}{\bf Sharpe Ratio}   & 1.1.-1.-1         & 60.3300           & 3.9800          & 29.5014          & 13.1852           \\
                                & 1.1.-1.30         & 60.3300           & 3.9800          & 29.5014          & 13.1852           \\
                                & 1.1.-1.14         & 60.3300           & 3.9800          & 29.5014          & 13.1852           \\
                                & 1.3.-1.-1         & 60.3300           & 3.9800          & 29.5014          & 13.1852           \\
                                & 1.3.-1.14         & 60.3300           & 3.9800          & 29.5014          & 13.1852      
                                \bigstrut\\ \hline
                                & \textbf{overall}  & \textbf{60.3300}  & \textbf{3.9800} & \textbf{29.5014} & \textbf{13.1852}  \\ 
\hline
\multirow{6}{*}{\bf Average Return} & 20.6.14.14        & 126.0000          & 14.0300         & 49.6879          & 29.3001           \\
                                & 20.6.14.30        & 121.8900          & 15.7400         & 49.3251          & 27.7299           \\
                                & 40.6.14.14        & 131.8100          & 9.7500          & 47.1009          & 30.4554           \\
                                & 40.6.14.30        & 141.2700          & 10.8900         & 50.9644          & 33.0127           \\
                                & 40.6.30.30        & 123.2500          & 10.6000         & 43.8679          & 32.6477           \\\hline
                                & \textbf{overall}  & \textbf{141.2700} & \textbf{9.7500} & \textbf{48.1893} & \textbf{30.7954} \bigstrut
\\ \hline
\end{tabular}
}
\caption{Top five parameters' relative returns (to buy-and-hold) with 10-minute trading period, chosen based on Sharpe Ratio and average returns.}
\label{fig:top-strategies}
\end{figure*}

\begin{figure*}[tb]
\centering\small 
\setlength{\tabcolsep}{4pt}
\begin{tabular}{|l|l|rrrr|} 
\hline
\textbf{Trading Period}       & \textbf{Parameter config.} & \textbf{Max.}         & \textbf{Min.}          & \textbf{Mean}           & \textbf{Std. dev.}  \bigstrut\\ 
\hline
\multirow{6}{*}{\textbf{1m}}  & 20.3.-1.-1       & 381.82          & -104.08          & 51.70          & 105.65              \bigstrut[t]\\
                     & 20.3.-1.14       & 281.35          & -102.42          & 63.40          & 86.36               \\
                     & 20.3.-1.30       & 281.35          & -102.42          & 63.34          & 86.37               \\
                     & 40.3.14.14       & 100.66          & 5.08             & 32.31          & 21.80               \\
                     & 40.3.14.30       & 102.52          & 5.08             & 32.64          & 22.35               \\ 
\cline{2-6}
                     & \textbf{overall} & \textbf{381.82} & \textbf{-104.08} & \textbf{48.68} & \textbf{74.86}    \bigstrut   \\ 
\hline
\multirow{6}{*}{\textbf{10m}} & 20.6.14.14       & 126.00          & 14.03            & 49.69          & 29.30               \bigstrut[t]\\
                     & 20.6.14.30       & 121.89          & 15.74            & 49.33          & 27.73               \\
                     & 40.6.14.14       & 131.81          & 9.75             & 47.10          & 30.46               \\
                     & 40.6.14.30       & 141.27          & 10.89            & 50.96          & 33.01               \\
                     & 40.6.30.30       & 123.25          & 10.60            & 43.87          & 32.65               \\ 
\cline{2-6}
                     & \textbf{overall} & \textbf{141.27} & \textbf{9.75}    & \textbf{48.19} & \textbf{30.80}      \\ 
\hline
\multirow{6}{*}{\textbf{1h}}  & 40.1.14.-1       & 130.69          & 9.44             & 48.84          & 30.98               \bigstrut[t]\\
                     & 40.1.14.14       & 131.66          & 9.44             & 49.05          & 31.30               \\
                     & 40.1.14.30       & 131.66          & 9.44             & 49.05          & 31.30               \\
                     & 20.3.30.30       & 126.21          & 0.23             & 40.55          & 30.95               \\
                     & 40.6.30.30       & 117.89          & 14.28            & 47.38          & 25.66               \\ 
\cline{2-6}
                     & \textbf{overall} & \textbf{131.66} & \textbf{0.23}    & \textbf{46.97} & \textbf{30.30}      
                     \bigstrut\\
\hline
\end{tabular}
\caption{Comparison of relative returns (to buy-and-hold) for the top 5 strategies of 1-hour, 10-minute and 1-minute trading periods. Top strategies selected based on average returns.}
\label{fig:frequency}
\end{figure*}

\section{Experimental Analysis}
\label{sec:experimental}

In this section, we present our evaluation of algorithmic training and our \tool implementation's end-to-end execution. The results of the former suggest that parametrising and pre-training trading algorithms on historical data does improve returns significantly, highlighting the value of hiding some parameters for achieving algorithmic privacy. 
The second major concern in addition to profitability is whether on-chain trading can be fast and cost-effective enough. 
Our time and gas costs of a single round of \tool,~\MeanEndToEndExecutionTime and~\GasUsedTotal, respectively, are acceptable in practice as we point out that there is no need to trade, for instance, at every block. We demonstrate that trades do not have to be placed so frequently without much loss of revenue. 

\subsection{Algorithmic Training}
\label{sec:eval-algorithmic-training}
\point{Setup}
Algorithmic training requires an environment for implementing and running the trading algorithm on historical data. We use Zenbot\footnote{\url{https://github.com/DeviaVir/zenbot}}, an open-source off-chain cryptocurrency trading bot with backtesting features. We use historical ETH-USDC price between~25.08.2020 and~25.08.2021, retrieved from the Poloniex exchange.

\point{Training}
We perform grid search through the training periods and within the space of minimum, maximum, and mean values that the parameters described in Section \ref{sec:overview-algo-training} can take, with the limiting values for each parameter being: $1\leq\mathit{M}\leq40$, $1\leq\mathit{D}\leq6$, $-1\leq\mathit{U}\leq30$, $-1\leq\mathit{L}\leq30$. Once we record the algorithm's performances using all parameter configurations in this space, the next step is determining the top-performing parameters.

Since many factors such as profitability and risk-aversion can be of varying importance in different settings, there is not a single method for determining a trading algorithm's success. Therefore, we use two different approaches to score the parameter configurations based on their performances: average return and \emph{Sharpe ratio}. Sharpe ratio, otherwise known as \emph{reward-to-variability ratio} is a well-known measure used in technical portfolio management, characterised by the difference between the average return of an investment strategy and a riskless investment, divided by the standard deviation of the former~\cite{sharpe1994sharpe}:

\begin{gather}
S_a = \frac{E[R_a-R_b]}{\sigma_a} = \frac{E[R_a-R_b]}{\sqrt{\mathrm{var}[R_a-R_b]}}
\end{gather}

\point{Testing} 
As Figure~\ref{fig:top-strategies} shows, the top-ranking parameters based  on average
\begin{wrapfigure}{r}{0.5\textwidth}
\centering
\includegraphics[width=0.5\textwidth]{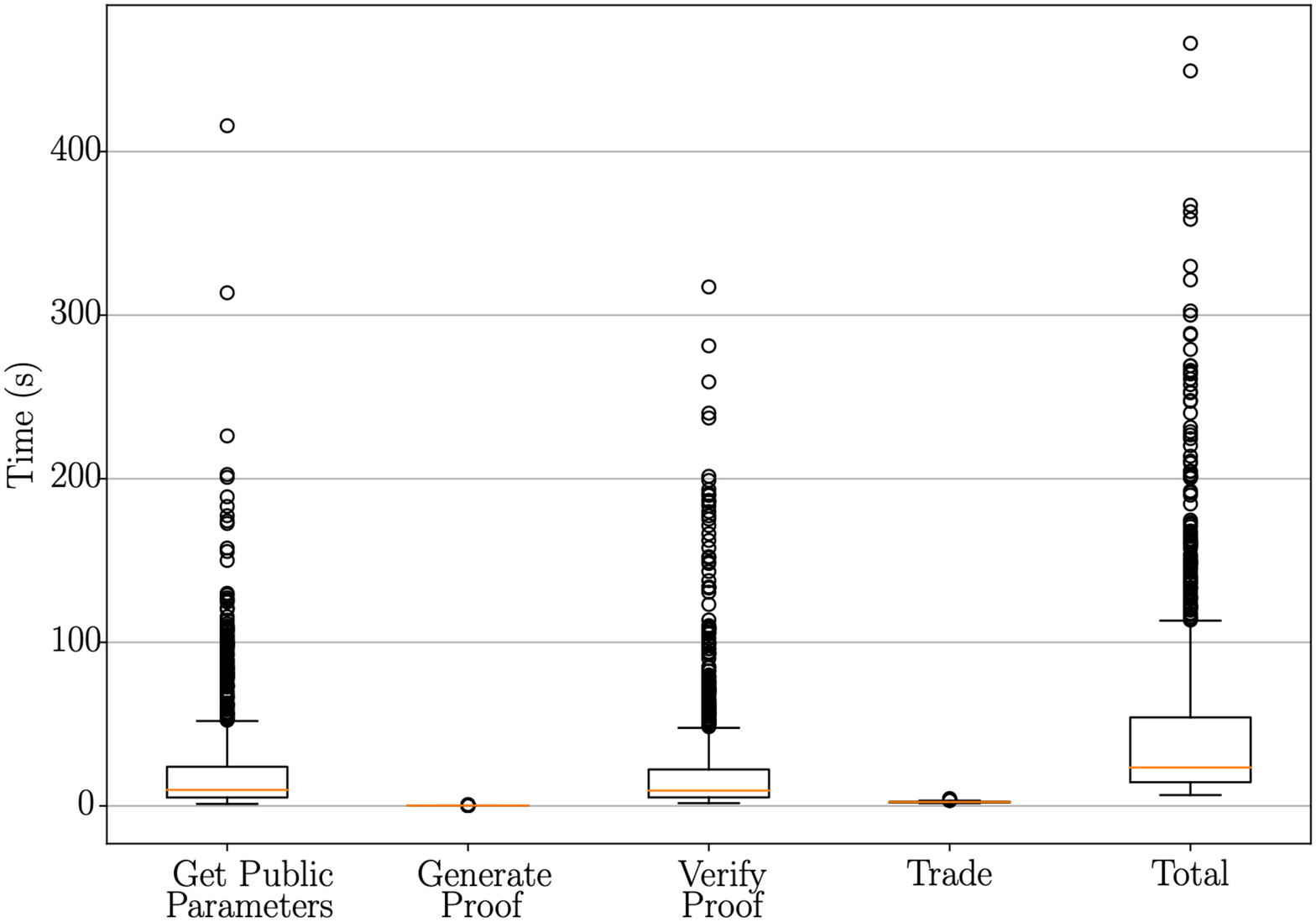}
\caption{Boxplot showing time performances of (a) public parameter retrieval, (b) proof generation, (c) proof verification, (d) trading, (e) end-to-end run over 1000 trading instances.}
\label{fig:performance}
\end{wrapfigure}
 return made up to~141\% more return compared to buy-and-hold. 

Although this value remained at~$60\%$ for all configurations of Sharpe ratio, this ranking strategy performed with significantly lower standard deviation: overall less than half of average return, at~$13\%$. 

These results are in-line with the characteristics of two ranking approaches: Average return strategy aims to maximise returns, while Sharpe ratio additionally takes risk associated with the return into account. It is also important to note that all parameter configurations performed better than the buy-and-hold strategy in all testing periods.

\point{Summary} Our evaluation of algorithmic training demonstrates that even with a simplistic trading algorithm and training method, parametrization and training can significantly improve trading performance. This finding suggests that parameter privacy can be an essential component of achieving algorithmic privacy.

\subsection{End-to-End Performance}
\point{Time measurements}
We evaluate \tool implementation's performance through~$1000$ end-to-end executions, as captured in Figure~\ref{fig:performance}, which involves 
\begin{enumerate}
\item  retrieving and calculating public parameters on-chain, 
\item  generating the proof of trading decision, 
\item  verifying it on-chain, and 
\item  trading on DeversiFi. 
\end{enumerate}
The experiments were performed on a local machine (2-core Intel~i5-5257U~CPU clocked at 2,7~GHz, with~8GB~RAM,~256GB~SSD), which is suitable for our evaluation as the off-chain computations are not computationally intensive -- The results show, proof generation, which is the only calculation executed off-chain and locally, takes the least amount of time among all operations, around~$0.2$ to~$0.8$ seconds. The time cost of end-to-end execution, is~\MeanEndToEndExecutionTime on average.

Although the time needed for the two on-chain tasks constitute the substantial amount of end-to-end execution time vary significantly (presumably affected by the network conditions), it invariably constitutes the substantial amount of end-to-end execution time. It takes between 1.2~and~415.8, on average~22.8 seconds to get public parameters, i.e. retrieve price from the oracle and calculate the Bollinger bands. Time required to verify the off-chain bot's trading decision zero-knowledge proof is~1.6~to~317.2 seconds, averaging at 23~seconds. Once all the necessary steps are complete, it takes approximately~2.3 seconds to execute the actual trade on DeversiFi. 

\point{Gas costs} 
Interactions with three entities cost gas in \tool: the on-chain bot, the verifier contract and the on-chain price oracle. Gas used for on-chain bot including the gas spent on retrieving price from the on-chain oracle is steady at~\GasUsedForPublicParams. Unlike these two components which perform the same computation on each call, the verifier contract's calculations, and therefore gas usage slightly varies among calls, with an average of~\GasUsedForVerification. As a result, the average gas cost of end-to-end execution is around~0,04592ETH~=~$\sim$\$150USD (as of~28.08.2021).

\point{Frequency of execution} 
We investigate the trading frequency's effect on returns, motivated by minimizing the aggravated execution costs, since gas costs are intrinsically (also demonstrated by our results) almost constant per trade. We train and test the Bollinger algorithm as described in Section~\ref{sec:overview-algo-training} with~1-minute,~10-minute, and~1-hour trading periods. As Figure~\ref{fig:frequency} shows, although there is an inverse relationship between trading period length and mean relative returns, the difference between~1-hour and~1-minute trading periods is less than~2\%. Moreover, the results indicate that lower trading periods may lead to higher standard deviation. Overall, trading less often might make similar returns with lower risk, therefore can be a useful strategy for reducing gas costs. 
\point{Summary} 
The average \MeanEndToEndExecutionTime time cost of end-to-end execution is dominated by the on-chain operations, which show highly variable time performance and cost around~\GasUsedTotal. As frequency of execution does not effect rate of return significantly, trading with lower frequencies can yield similar profit as well as reducing gas costs.

\section{Limitations and Future Work}
\label{sec:future}

\point{Slippage}
An asset's price change in the course of a trade, namely \emph{slippage}, and it might cause discrepancy between the trading algorithm's expected and realized trade decisions. Slippage's effect on trades is algorithm-dependent, since an algorithm may compare the price at the moment of decision to various values derived from that price. Moreover, an algorithm can use past prices, as in the case of Bollinger algorithm, using indicators such as Simple Moving Average (SMA) or Time-Weighted Average Price (TWAP), likely increasing susceptibility to being affected by slippage. Further research is needed to provide a systematic understanding these effects.

\point{Off-Chain Integrity Concerns} Currently in StarkEx and DeversiFi, the only way of verifying that a specific transaction was included in a proof batch is to have access to the sequencer data. Therefore, users have no direct way of verifying that the off-chain bot correctly executes the verified trade decisions. This is not a shortcoming of our design but the systems available. To mitigate this issue of the existing systems, a supervising committee can be appointed to observe off-chain bot's trade decisions and executions, making a trade-off between making confidential trade information available to the committee members and establishing end-to-end integrity. 

\point{Programming Language} \tool development requires implementing and integrating multiple programs with various programming languages and frameworks, making the process intricate and potentially fragile. A natural progression of this work is to create a programming language that is expressive enough to cover the use cases and be compiled/transpiled to the necessary languages. Such a language may noticeably increase the robustness and practicality of deployment and maintenance of the bot, as well as preventing data leaks with type-checking. Although systems such as Hawk~\cite{Kosba2016}, Arbitrum~\cite{Kalodner2018}, Ekiden~\cite{Cheng2019}, and {zkay}~\cite{Report2020,Steffen2019} language aim to achieve easy development of private blockchain programming, none of them quite provide the components required to develop a \tool.

\section{Conclusion}
\label{sec:conclusion}

The aim of this research was to present a system for automating trading on blockchains in a way that promotes minimized trust without losing the competitive advantage. We introduce a hybrid approach, partitioning the trading process into on- and off-chain components. Our approach uses zero-knowledge proofs to maintain end-to-end integrity and privacy-enhancing DEXs to prevent trades from being mirrored and private parameters from being reverse engineered. 

We have presented a sample trading bot using our strategy and a simple trading algorithm that uses Bollinger Bands. Parametrizing, training and testing our bot on historical data has resulted in around~$\AvgMultipleFromTrading\times$ and up to~$\MultipleFromTrading\times$ returns relative to the baseline buy-and-hold strategy.
In our implementation, an average user willing to invest~\$\UserInvestment with \tool would earn~\$\MonthlyEarning, while spending~\$\MonthlyGasCost on gas; this assumes a user pool of~1,000 subscribers. 

Our measurements show that the end-to-end average execution time for our implemented system is~\MeanEndToEndExecutionTime. We demonstrate that lowering the trading frequency does not significantly affect the performance of the algorithm. This result also indicates that by trading less often, gas costs can be reduced without sacrificing returns. These observation show that \tool is a viable and cost-effective strategy for building on-chain trading systems.

\newpage\appendix
\section{Implementation Details}
\label{sec:impl}
This section describes the implementation details of the system outlined in Section~\ref{sec:overview}.
We cover algorithmic training in Appendix~\ref{sec:appendix-training}, Zokrates details in Appendix~\ref{sec:zokrates}, and DiversiFi exchange interactions in Appendix~\ref{sec:diversifi}. 

\subsection{Algorithmic Training}
\label{sec:appendix-training}
Our algorithmic training is essentially a search function, which looks for the top-performing parameter configurations in a past training period. These configurations are then tested for their ex-post performances in a testing period.

\point{Bollinger Bands}
Below is the formula for Bollinger Bands, with a simple moving average (of size $N$) using a number ($d$) of standard deviations \cite{bollinger1992using}

\begin{equation}
\begin{gathered}
\overline{X} \ =\ \ \frac{\sum _{j\ =\ 1}^{N} X_{j}}{N} \\
\sigma \ =\ \sqrt{\frac{\sum _{j\ =\ 1}^{N}( X_{j} -\overline{X})^{2}}{N}} \label{eq:bollinger}\\
\mathit{Upper\ BB}\ =\ \overline{X} \ +\ d\ *\ \sigma  \\
\mathit{Lower\ BB}\ =\overline{X} \ -\ d\ *\ \sigma \\
\mathit{where\ X\ is\ closing\ price}
\end{gathered}
\end{equation}
Our trading algorithm can be formulized as:

\begin{equation}
\begin{gathered}
\mathit{buy}:\ X\ < (\mathit{Lower\ BB}\ /\ 100) \ *\ ( 100\ +\ L) \label{eq:trading-algo}\\
\mathit{sell}:\ X\  > (\mathit{Upper\ BB}\ /\ 100) \ *\ ( 100\ -\ U)
\end{gathered}
\end{equation}

\point{Training}
Using grid search in the space of the minimum, maximum, and mean values that the parameters can take, results in a set \emph{C}, containing all parameter configurations obtained by the product of the sets of possible values for each parameter:

\begin{equation}
    \begin{gathered}
N'=f(N) \land D'=f(D) \land L'=f(L) \land U'=f(U) \label{eq:product}\\
C= N'\times D'\times L'\times U'\\
where \ f(A)=\{min\ A,\ max\ A,\ mean\ A\}
\end{gathered}
\end{equation}

\noindent
Running the trading algorithm with all elements of $\mathit{C}$ through the training periods, we rank the parameter configurations based on their performance.

\begin{figure}[tb]
    \begin{lstlisting}[basicstyle=\small, breaklines=true,  , basicstyle=\ttfamily, postbreak=\mbox{$\hookrightarrow$\space}]
def main(field currentPrice, 
    field upperBollingerBand, 
    field lowerBollingerBand, 
    private field buy_sell_flag, 
    private field boundPercentage) -> bool:
    // buy_sell_flag is 1 for buy, 0 for sell
    bool verified = if buy_sell_flag == 1 then currentPrice < (lowerBollingerBand / 100) * (100 + boundPercentage) else currentPrice > (upperBollingerBand / 100) * (100 - boundPercentage) fi
      return verified
    \end{lstlisting}
    \caption{Zokrates code in our implementation for generating verifier smart contracts and proofs.}
    \label{fig:zokrates-code}
\end{figure}

\subsection{Zokrates Interactions}
\label{sec:zokrates}

\point{The Zokrates program}
Zokrates~\cite{eberhardt2018zokrates} toolbox allows writing a root program in Zokrates language, which specifies the knowledge to be proven and acts as the originator for a verifier smart contract as well as a proof generator. The Zokrates program we have implemented for verifying the trade decisions of the Bollinger algorithm is presented in Figure~\ref{fig:zokrates-code}. The program takes five parameters: (1) $X$ (the current price), (2) $\mathit{Upper\ BB}$, (3) $\mathit{Lower\ BB}$, (4) a flag that tells the code to generate a proof for a buy or sell decision, and lastly, (5) either $\mathit{L}$ or $\mathit{U}$ based on the decision type. Logically, the program checks if the inputted $X$, $\mathit{Upper\ BB}$, $\mathit{Lower\ BB}$, $\mathit{L}$ or $\mathit{U}$ values lead to the trading decision denoted by the flag. 
A proof generated from this root program confirms that the prover has made an undisclosed trading decision based on the Bollinger algorithm, using some hidden parameters and the public parameters it has obtained from the on-chain bot.

\point{Verifying public parameters}
Correctness of a proof ensures that a trading decision is made accurately with the parameters inputted to the proof generator, but it does not guarantee that the inputted parameters are the correct ones. In other words, it is possible to produce a verifiable proof using parameter values that are different than the on-chain bot or the algorithmic trainer provides, as long as the trading algorithm is executed correctly. Although the user needs to trust the \tool with the correctness and integrity of private parameters, it is necessary to implement a mechanism for ensuring that the correct public parameters are used in proof generation.

We achieve this through utilizing a Zokrates feature: The inputs to the root Zokrates program can be public or private; the former are included in the proof as plaintext, while the latter are not revealed. This enables the off-chain bot to feed the price and other public trading parameters into the proof as public parameters and the on-chain bot to check if the values included in the proof are the ones it has provided. This simple confirmation is enough to accomplish the desired task: if the off-chain bot uses incorrect public parameters, this will be visible in the proof, and the proof verification will fail. 

\point{Preventing information leaks}
As mentioned in Section~\ref{sec:private-trading}, using zero-knowledge proofs to keep some parameters private is not sufficient for achieving algorithmic privacy; whether the bot has decided to buy or sell should also be kept secret. Unless designed properly, on-chain proof verification can leak this information. There are two conditions that the verification mechanism should hold to prevent leaking the trade decision:

\begin{enumerate}
    \item 
There must be a \emph{single} proof generator and verifier contract for all possible conditions that lead to a trade decision. Otherwise, the trading decision is visible through the type of contract that the off-chain bot commands the on-chain bot to relay the proof to. 
    \item 
Trade decision should be specified as a private input in the root Zokrates code to prevent it from being openly included in the proof.
\end{enumerate}

\point{Validating the verifier contract} The users should be able to ensure that the verifier contract is verifying what it is supposed to be verifying. The Zokrates verification system is based on Arithmetic Circuits and Rank-1-Constraint-Systems (R1CS)~\cite{eberhardt2018zokrates}. Consequently, even though the verifier contract is publicly visible, what it verifies is not easily understandable from the code. One way to enable verifier contract validation is making the root Zokrates code publicly accessible, e.g. by storing it in a smart contract. This way, the users can audit the high-level Zokrates code, compile it and compare the resulting Solidity code with the deployed smart contract. 

\subsection{DiversiFi Interactions}
\label{sec:diversifi}
Hiding the trade decisions of \tool is essential to protecting the privacy of its trading strategy. To execute the trades privately, our implementation uses DeversiFi~\cite{deversifi}, a self-custodial order book exchange that uses L2 for enhancing transaction privacy and speed. 

\point{Validium and transaction privacy}
DeversiFi is powered by the Validium-based StarkEx scalabilty engine~\cite{starkex}. Validium is an L2 solution where multiple transactions are batched off-chain in a single zero-knowledge proof, which is verified on-chain. Although Validium resembles \emph{ZK-rollups} in this sense, it differs in storing the entire data related to transactions off-chain.

Validium aims to solve the data availability problem by maintaining a \emph{Data Availability Committee~(DAC)}, where the (decentralized) members always hold a copy of all off-chain transaction data and sign a commitment to every new state before it can be submitted on-chain. Limiting the access to transaction data to the members of DAC, Validium enhances the trader's privacy, when compared to L1 schemes or other L2 solutions that keep transaction data on-chain. 

\point{Private trading on DeversiFi}
DeversiFi leverages Validium to allow its users to hide their trades from the public. User funds are stored in so-called \emph{token accounts} (or \emph{vaults}), which function as L2 wallets that can each have a unique ID, hold a single type of asset and can be assigned to a single user at a given time. A Merkle tree is formed from these token accounts and updated every time a token account's balance changes~\cite{howdeversifiworks}. 

Before starting to trade on DeversiFi, the user has to deposit funds into a StarkEx smart contract. The deposited amount is automatically recorded to the user's token account, making the funds available for trading. Trading happens entirely off-chain, with a maker-taker model: the maker signs an order, which includes the token types and amounts being traded, as well as the maker's account IDs for these tokens. The taker executes the trade by signing the order along with the IDs of their two corresponding token accounts. When a user wants to withdraw their funds from DeversiFi, the amount is subtracted from their token account balance and added to their ``on-chain withdrawal balance,'' enabling direct withdrawal from the StarkEx smart contract~\cite{howdeversifiworks,starkdex}.

\point{Limitations} Despite being a ``decentralised exchange'' in technical terms, DeversiFi does not provide a trustless trading platform. First, keeping trades private from other traders comes at the cost of sharing them with the DAC. The DAC members are well-known organisations which Ethereum blockchain users already share data with by using their services and infrastructures. In this sense, DeversiFi piggybacks on the existing implicit user trust in these organisations. 

Secondly, as it is DeversiFi who operates the \textit{sequencers}, which batch multiple transactions into a single zero-knowledge proof, the users have to trust DeversiFi with including their transaction in a proof and submitting it to the chain. In the worst scenario, if DeversiFi attempts to censor a user's a withdrawal request, the user can retrieve her funds from the StarkEx smart contract if she manages to acquire the necessary data from DAC members.

%
%
%
\bibliographystyle{splncs04}
\bibliography{bibliography}

\begin{thebibliography}{10}
\providecommand{\url}[1]{\texttt{#1}}
\providecommand{\urlprefix}{URL }
\providecommand{\doi}[1]{https://doi.org/#1}

\bibitem{3commas}
{3Commas}, \url{https://3commas.io/}, {Accessed: 2021-09-06}

\bibitem{coinmatics}
Coinmatics, \url{https://coinmatics.com/}, {Accessed: 2021-09-06}

\bibitem{deversifi}
{DeversiFi}, \url{https://www.deversifi.com/}, {Accessed: 2021-09-06}

\bibitem{howdeversifiworks}
How deversifi works,
  \url{https://docs.deversifi.com/articles#HowDeversiFiWorks}, {Accessed:
  2021-09-06}

\bibitem{shrimpy}
Shrimpy, \url{https://www.shrimpy.io/}, {Accessed: 2021-09-06}

\bibitem{starkex}
{StarkEx}: Powering scalable self-custodial transactions,
  \url{https://starkware.co/product/starkex/}, {Accessed: 2021-09-06}

\bibitem{tokensets}
{TokenSets}, \url{https://www.tokensets.com/}

\bibitem{tradelize}
Tradelize, \url{https://tradelize.com/}, {Accessed: 2021-09-06}

\bibitem{yearnfinance}
yearn.finance, \url{https://yearn.finance/}, {Accessed: 2021-09-06}

\bibitem{zcash-ceremony}
Zcash parameter generation, \url{https://z.cash/technology/paramgen/},
  {Accessed: 2021-09-06}

\bibitem{Adams2021}
Adams, H., Zinsmeister, N., Salem~moody, M., Keefer, R., Robinson, D.: Uniswap
  v3 core. Whitepaper  (March 2021),
  \url{https://uniswap.org/whitepaper-v3.pdf}

\bibitem{Banerjee2021}
Banerjee, A., Clear, M., Tewari, H.: {zkHawk}: Practical private smart
  contracts from {MPC}-based {Hawk}. arXiv preprint arXiv:2104.09180  (2021)

\bibitem{Report2020}
Baumann, N., Steffen, S., Bichsel, B., Tsankov, P., Vechev, M.: zkay v0.2:
  Practical data privacy for smart contracts. arXiv preprint arXiv:2009.01020
  (2020)

\bibitem{beres2020blockchain}
B{\'e}res, F., Seres, I.A., Bencz{\'u}r, A.A., Quintyne-Collins, M.: Blockchain
  is watching you: Profiling and deanonymizing ethereum users. arXiv preprint
  arXiv:2005.14051  (2020)

\bibitem{bollinger1992using}
Bollinger, J.: Using bollinger bands. Stocks \& Commodities  (1992)

\bibitem{Bowe2020}
Bowe, S., Chiesa, A., Green, M., Miers, I., Mishra, P., Wu, H.: {ZEXE}:
  Enabling decentralized private computation. In: IEEE Symposium on Security
  and Privacy (SP) (2020). \doi{10.1109/SP40000.2020.00050}

\bibitem{bunz2020zether}
B{\"u}nz, B., Agrawal, S., Zamani, M., Boneh, D.: Zether: Towards privacy in a
  smart contract world. In: International Conference on Financial Cryptography
  and Data Security. Springer (2020). \doi{10.1007/978-3-030-51280-4\_23}

\bibitem{rollups}
Buterin, V.: An incomplete guide to rollups (May 2021),
  \url{https://vitalik.ca/general/2021/01/05/rollup.html}

\bibitem{Calcaterra2021}
Calcaterra, C., Kaal, W.A.: {Decentralized Finance (DeFi)}. SSRN Electronic
  Journal  (2021). \doi{10.2139/ssrn.3782216}

\bibitem{cecchetti2017solidus}
Cecchetti, E., Zhang, F., Ji, Y., Kosba, A., Juels, A., Shi, E.: Solidus:
  Confidential distributed ledger transactions via {PVORM}. In: ACM SIGSAC
  Conference on Computer and Communications Security (2017).
  \doi{10.1145/3133956.3134010}

\bibitem{chan2021quantitative}
Chan, E.P.: Quantitative trading: how to build your own algorithmic trading
  business. John Wiley \& Sons (2021)

\bibitem{chan2013algorithmic}
Chan, E.P.: Algorithmic trading: winning strategies and their rationale. John
  Wiley \& Sons (2013)

\bibitem{Cheng2019}
Cheng, R., Zhang, F., Kos, J., He, W., Hynes, N., Johnson, N., Juels, A.,
  Miller, A., Song, D.: Ekiden: A platform for confidentiality-preserving,
  trustworthy, and performant smart contracts. In: IEEE European Symposium on
  Security and Privacy (EuroS\&P) (2019). \doi{10.1109/EuroSP.2019.00023}

\bibitem{chervinski2019floodxmr}
Chervinski, J.O.M., Kreutz, D., Yu, J.: Floodxmr: Low-cost transaction flooding
  attack with monero's bulletproof protocol. IACR Cryptology ePrint Archive
  (2019), \url{https://eprint.iacr.org/2019/455.pdf}

\bibitem{danezis2013pinocchio}
Danezis, G., Fournet, C., Kohlweiss, M., Parno, B.: Pinocchio coin: building
  {Zerocoin} from a succinct pairing-based proof system. In: The First {ACM}
  workshop on Language support for privacy-enhancing technologies (2013).
  \doi{10.1145/2517872.2517878}

\bibitem{eberhardt2018zokrates}
Eberhardt, J., Tai, S.: Zokrates - scalable privacy-preserving off-chain
  computations. In: IEEE International Conference on Internet of Things
  (iThings) and IEEE Green Computing and Communications (GreenCom) and IEEE
  Cyber, Physical and Social Computing (CPSCom) and IEEE Smart Data
  (SmartData). IEEE (2018). \doi{10.1109/Cybermatics\_2018.2018.00199}

\bibitem{Ellis2017}
Ellis, S., Juels, A., Nazarov, S.: {ChainLink}: A decentralized oracle network
  (2017), \url{https://link.smartcontract.com/whitepaper}

\bibitem{eskandari2021sok}
Eskandari, S., Salehi, M., Gu, W.C., Clark, J.: Sok: Oracles from the ground
  truth to market manipulation. arXiv preprint arXiv:2106.00667  (2021)

\bibitem{fauzi2019quisquis}
Fauzi, P., Meiklejohn, S., Mercer, R., Orlandi, C.: Quisquis: A new design for
  anonymous cryptocurrencies. In: International Conference on the Theory and
  Application of Cryptology and Information Security (2019).
  \doi{10.1007/978-3-030-34578-5\_23}

\bibitem{gervais2016security}
Gervais, A., Karame, G.O., W{\"u}st, K., Glykantzis, V., Ritzdorf, H., Capkun,
  S.: On the security and performance of proof of work blockchains. In:
  Proceedings of the 2016 ACM SIGSAC conference on computer and communications
  security. pp. 3--16 (2016)

\bibitem{gudgeon2020sok}
Gudgeon, L., Moreno-Sanchez, P., Roos, S., McCorry, P., Gervais, A.: Sok:
  Layer-two blockchain protocols. In: International Conference on Financial
  Cryptography and Data Security. Springer (2020).
  \doi{10.1007/978-3-030-51280-4\_12}

\bibitem{hendershott2009algorithmic}
Hendershott, T., Riordan, R., et~al.: Algorithmic trading and information.
  Manuscript, University of California, Berkeley  (2009).
  \doi{10.2139/ssrn.1472050}

\bibitem{hopwood2016zcash}
Hopwood, D., Bowe, S., Hornby, T., Wilcox, N.: Zcash protocol specification.
  GitHub: San Francisco, CA, USA  (2016),
  \url{https://zips.z.cash/protocol/protocol.pdf}

\bibitem{jiang2017cryptocurrency}
Jiang, Z., Liang, J.: Cryptocurrency portfolio management with deep
  reinforcement learning. In: IEEE Intelligent Systems Conference (IntelliSys)
  (2017). \doi{10.1109/IntelliSys.2017.8324237}

\bibitem{jivanyan2019lelantus}
Jivanyan, A.: Lelantus: Towards confidentiality and anonymity of blockchain
  transactions from standard assumptions. {IACR} Cryptology {ePrint} Archive
  (2019), \url{https://eprint.iacr.org/2019/373/20190604:053917}

\bibitem{Kalodner2018}
Kalodner, H., Goldfeder, S., Chen, X., Weinberg, S.M., Felten, E.W.: Arbitrum:
  Scalable, private smart contracts. In: 27th {USENIX} Security Symposium
  (2018). \doi{10.5555/3277203.3277305}

\bibitem{kim2010electronic}
Kim, K.: Electronic and algorithmic trading technology: the complete guide.
  Academic Press (2010)

\bibitem{Kosba2016}
Kosba, A., Miller, A., Shi, E., Wen, Z., Papamanthou, C.: {Hawk: The Blockchain
  Model of Cryptography and Privacy-Preserving Smart Contracts}. IEEE Symposium
  on Security and Privacy (SP)  (2016). \doi{10.1109/SP.2016.55}

\bibitem{le2020amr}
Le, D.V., Gervais, A.: Amr: Autonomous coin mixer with privacy preserving
  reward distribution. ACM Conference on Advances in Financial Technologies
  (AFT’21)  (2021)

\bibitem{Maharramov2019}
Maharramov, T., Francioni, E.: The {Dusk} network whitepaper

\bibitem{makarov2020trading}
Makarov, I., Schoar, A.: Trading and arbitrage in cryptocurrency markets.
  Journal of Financial Economics  (2020). \doi{10.2139/ssrn.3171204}

\bibitem{monero}
Noether, S.: Ring signature confidential transactions for {Monero}. IACR
  Cryptology ePrint Archive (2015), \url{https://eprint.iacr.org/2015/1098}

\bibitem{pole2011statistical}
Pole, A.: Statistical arbitrage: algorithmic trading insights and techniques.
  John Wiley \& Sons (2011)

\bibitem{poon2017plasma}
Poon, J., Buterin, V.: Plasma: Scalable autonomous smart contracts. Whitepaper
  (2017)

\bibitem{poon2016lightning}
Poon, J., Dryja, T.: The bitcoin lightning network: Scalable off-chain instant
  payments (2016), \url{https://lightning.network/lightning-network-paper.pdf}

\bibitem{qin2021cefi}
Qin, K., Zhou, L., Afonin, Y., Lazzaretti, L., Gervais, A.: Cefi vs.
  defi--comparing centralized to decentralized finance. arXiv preprint
  arXiv:2106.08157  (2021)

\bibitem{qin2021quantifying}
Qin, K., Zhou, L., Gervais, A.: Quantifying blockchain extractable value: How
  dark is the forest? arXiv preprint arXiv:2101.05511  (2021)

\bibitem{qin2020attacking}
Qin, K., Zhou, L., Livshits, B., Gervais, A.: Attacking the defi ecosystem with
  flash loans for fun and profit. Financial Cryptography and Data Security:
  25th International Conference (FC'21)  (2021)

\bibitem{Sanchez2018}
S{\'a}nchez, D.C.: Raziel: Private and verifiable smart contracts on
  blockchains. arXiv preprint arXiv:1807.09484  (2018)

\bibitem{sasson2014zerocash}
Sasson, E.B., Chiesa, A., Garman, C., Green, M., Miers, I., Tromer, E., Virza,
  M.: Zerocash: Decentralized anonymous payments from {Bitcoin}. In: IEEE
  Symposium on Security and Privacy (SP) (2014). \doi{10.1109/SP.2014.36}

\bibitem{shahandzhang}
Shah, D., Zhang, K.: Bayesian regression and bitcoin. In: 52nd Annual Allerton
  Conference on Communication, Control, and Computing (Allerton) (2014).
  \doi{10.1109/ALLERTON.2014.7028484}

\bibitem{sharpe1994sharpe}
Sharpe, W.F.: The sharpe ratio. Journal of portfolio management  (1994)

\bibitem{starkdex}
StarkWare: Starkdex deep dive: Introduction (2021),
  \url{https://medium.com/starkware/starkdex-deep-dive-introduction-7b4ef0dedba8}

\bibitem{Steffen2019}
Steffen, S., Bichsel, B., Gersbach, M., Melchior, N., Tsankov, P., Vechev, M.:
  zkay: Specifying and enforcing data privacy in smart contracts. In: ACM
  SIGSAC Conference on Computer and Communications Security (2019).
  \doi{10.1145/3319535.3363222}

\bibitem{Teutsch2019}
Teutsch, J., Reitwie{\ss}ner, C.: A scalable verification solution for
  blockchains. arXiv preprint arXiv:1908.04756  (2019)

\bibitem{treleaven2013algorithmic}
Treleaven, P., Galas, M., Lalchand, V.: Algorithmic trading review.
  Communications of the ACM  (2013). \doi{10.1145/2500117}

\bibitem{van2013cryptonote}
Van~Saberhagen, N.: Cryptonote v 2.0 (2013),
  \url{https://www.getmonero.org/ru/resources/research-lab/pubs/whitepaper_annotated.pdf}

\bibitem{vo2020high}
Vo, A., Yost-Bremm, C.: A high-frequency algorithmic trading strategy for
  cryptocurrency. Journal of Computer Information Systems  (2020).
  \doi{10.1080/08874417.2018.1552090}

\bibitem{moneroAttack}
Wijaya, D.A., Liu, J., Steinfeld, R., Liu, D.: Monero ring attack: Recreating
  zero mixin transaction effect. In: 17th IEEE International Conference On
  Trust, Security And Privacy In Computing And Communications/ 12th IEEE
  International Conference On Big Data Science And Engineering
  (TrustCom/BigDataSE) (2018). \doi{10.1109/TrustCom/BigDataSE.2018.00165}

\bibitem{binancereuters}
Wilson, T.: Explainer: Binance: The crypto giant facing pressure from
  regulators. Reuters (08 2021),
  \url{https://www.reuters.com/business/finance/binance-crypto-giant-facing-pressure-regulators-2021-08-19/}

\bibitem{tracingYoussaf}
Yousaf, H., Kappos, G., Meiklejohn, S.: Tracing transactions across
  cryptocurrency ledgers. In: 28th {USENIX} Conference on Security Symposium
  (2019). \doi{10.5555/3361338.3361396}

\bibitem{zhang2019security}
Zhang, R., Xue, R., Liu, L.: Security and privacy on blockchain. ACM Computing
  Surveys (CSUR)  (2019). \doi{10.1145/3316481}

\bibitem{zhou2021just}
Zhou, L., Qin, K., Cully, A., Livshits, B., Gervais, A.: On the just-in-time
  discovery of profit-generating transactions in {DeFi} protocols. IEEE
  Symposium on Security and Privacy (SP'21)  (2021)

\bibitem{zhou2021a2mm}
Zhou, L., Qin, K., Gervais, A.: {A2MM}: Mitigating frontrunning, transaction
  reordering and consensus instability in decentralized exchanges. arXiv
  preprint arXiv:2106.07371  (2021)

\bibitem{Zhou2020}
Zhou, L., Qin, K., Torres, C.F., Le, D.V., Gervais, A.: High-frequency trading
  on decentralized on-chain exchanges. In: IEEE Symposium on Security and
  Privacy (SP) (2021), \url{https://arxiv.org/pdf/2009.14021.pdf}

\bibitem{Zyskind2019}
Zyskind, G.: {Enigma: Decentralized Computation Platform with Guaranteed
  Privacy}. New Solutions for Cybersecurity  (2019).
  \doi{10.7551/mitpress/11636.003.0018}

\end{thebibliography}

\end{document}